\documentclass{svproc}
\usepackage{url}

\usepackage{amsfonts}
\usepackage{xparse}
\usepackage{amsmath}
\usepackage{amssymb}
\usepackage{mathtools}
\usepackage{tcolorbox}
\usepackage{romannum}
\usepackage{pdfpages}
\usepackage{fancyhdr}
\newcommand{\eg}{{\it e.g.,}\ }
\newcommand{\ie}{{\it i.e.,}\ }

\newcommand{\mt}[1]{\textrm{\tiny #1}}
\newcommand{\reef}[1]{(\ref{#1})}

\renewcommand{\(}{\left(}
\renewcommand{\)}{\right)}
\renewcommand{\[}{\left[}
\renewcommand{\]}{\right]}

\newcommand{\GN}{G_\mt{N}}

\newcommand{\mC}{\mathcal{C}}

\newcommand{\be}{\begin{equation}}
	\newcommand{\ee}{\end{equation}}
\newcommand{\bea}{\begin{eqnarray}}
	\newcommand{\eea}{\end{eqnarray}}
\newcommand{\ba}{\begin{align}}
	\newcommand{\ea}{\end{align}}
\newcommand{\beq}{\begin{equation}}
	\newcommand{\eeq}{\end{equation}}
\newcommand{\beqa}{\begin{eqnarray}}
	\newcommand{\eeqa}{\end{eqnarray}}

\newcommand{\cv}{{\cal C}_\mt{V}}
\newcommand{\ca}{{\cal C}_\mt{A}}
\newcommand{\cvv}{{\cal C}_\mt{SV}}
\newcommand{\Scft}{\Sigma_\mt{CFT}}

\begin{document}	
	\mainmatter              
	\title{Complexity Equals (Almost) Anything}
	\titlerunning{Complexity Equals (Almost) Anything}  
	%
	\author{Robert C. Myers\inst{1} \and Shan-Ming Ruan\inst{2}}
	\authorrunning{Myers and Ruan} 
	%
	%
	\institute{Perimeter Institute for Theoretical Physics, \\
		Waterloo, ON N2L 2Y5, Canada\\
		\email{rmyers@perimeterinstitute.ca},
		\and
		Center for Gravitational Physics and Quantum Information, \\
		Yukawa Institute for Theoretical Physics, Kyoto University,\\
		Kitashirakawa Oiwakecho, Sakyo-ku, Kyoto 606-8502, Japan\\
		\email{ruan.shanming@yukawa.kyoto-u.ac.jp}}

	\maketitle              

\begin{abstract}
Recent investigations \cite{Belin:2021bga,Belin:2022xmt,Jorstad:2023kmq} have introduced an infinite class of novel gravitational observables in Asymptotically anti-de Sitter (AdS) space that reside on codimension-one or -zero regions of the bulk spacetime. These observables encompass well-established holographic complexity measures such as the maximum volume of the extremal hypersurfaces and the action or spacetime volume of the Wheeler-DeWitt (WDW) patch. Furthermore, this family of observables exhibits two universal properties when applied to the thermofield double (TFD) state: they exhibit linear growth at late times and faithfully reproduce the switchback effect. This implies that any observable from this class has the potential to serve as a gravitational dual for the circuit complexity of boundary states.
		\\
	   {{\bf{Report Number:}} YITP-24-34}
\end{abstract}

\section{Why Holographic Complexity?}
In recent years, there has been significant progress in understanding quantum gravity from the perspective of quantum information theory, especially in the context of holography. Quantum information theory has provided fascinating and profound insights into fundamental questions about the AdS/CFT correspondence, a powerful framework for studying the holographic duality between bulk gravitational theory and quantum field theory on the boundary. In particular, the celebrated Ryu-Takayanagi formula proposed that the entanglement entropy of a subsystem of the holographic CFT is geometrically described by the area of a certain minimal surface in bulk spacetime \cite{Ryu:2006bv,Ryu:2006ef}. Developments in the past decade provide strong evidence that the entanglement structure of the underlying quantum mechanical degrees of freedom plays a crucial role in determining the emergent spacetime geometry and its dynamics.	
	
However, it has become clear that entanglement entropy alone is not sufficient to fully capture the time evolution of a quantum system. For example, it grows only for a short time until the system is thermalized. From a holographic perspective, this is reflected in the fact that holographic entanglement entropy is unable to probe the bulk of spacetime far beyond the event horizon of black holes. To overcome these limitations and to gain a more complete understanding of the underlying physical dynamics of the system, it is essential to investigate alternative quantum information measures. 
	
A fascinating  concept that has recently entered this discussion is the quantum circuit complexity \cite{Susskind:2014moa,Susskind:2014rva}. The latter  measures how difficult it is to create a particular target state $|\Psi_{\mt{T}} \rangle$ from a reference state $|\Psi_{\mt{R}} \rangle$ by using a set of fundamental unitary operations or ``gates''. For example, a quantum circuit can be realized by applying a set of elementary gates $g_i$:
	\begin{equation}
		|\Psi_{\mt{T}} \rangle = U_{\mt{TR}} |\Psi_{\mt{R}} \rangle = g_{n}\,g_{n-1}\,\cdots\, g_2\,g_{1}\, |\Psi_{\mt{R}} \rangle \,. 
	\end{equation} 
The circuit complexity of the target state $|\Psi_{\mt{T}} \rangle$ is then defined as the minimum number of gates needed to construct the particular unitary $U_{\mt{TR}}$, \ie the cost of the optimal circuit. Various approaches have also been developed in recent years to understand the complexity of states in quantum field theory, \eg see \cite{Jefferson:2017sdb,Chapman:2017rqy,Caputa:2017urj}. It is important to note that the definition of circuit complexity is ambiguous because the exact value of the complexity is sensitive to the choice of the reference state, the set of gates and the costs assigned to different types of gates. We emphasize that these ambiguities are inherent to the concept itself. 

A fascinating aspect of complexity is that it can continue to evolve long after the entanglement entropies of the quantum system have thermalized. In the following, we will focus on holographic complexity, \ie the bulk dual of the circuit complexity of the boundary states. Several proposals have been made to define holographic complexity. The complexity=volume (CV) proposal \cite{Susskind:2014rva,Stanford:2014jda} suggests that holographic complexity is dual to the maximal volume of a certain class of spacelike hypersurfaces, \ie 
	\begin{equation}\label{eq:defineCV}
		\cv(\Sigma_{\mt{CFT}}) =\,\max_{\Sigma_{\mt{CFT}}= \partial \Sigma}\left[\frac{\mathcal{V}(\Sigma)}{\GN \, \ell_{ \rm bulk}}\right] \,.
	\end{equation}
Here $\GN$ isthe Newton's constant of the bulk gravitational theory,
and $\ell_{ \rm bulk}$ is an arbitrary length scale, often chosen to be the AdS radius of curvature $L$. The codimension-one hypersurfaces $\Sigma$ are all anchored on the asymptotic boundary time slice $\Sigma_{\mt{CFT}}$ where the CFT state lives.  Different from the CV proposal, the complexity=action (CA) proposal \cite{Brown:2015bva} is associated with a codimension-zero subregion in the bulk spacetime. It proposes that holographic complexity is given by the gravitational action evaluated on the Wheeler-DeWitt patch, \ie the domain of dependence for the surfaces $\Sigma$ in eq.~\reef{eq:defineCV},
	\begin{equation}\label{eq:defineCA}
	\ca(\Sigma_{\mt{CFT}}) =  \frac{I_\mt{WDW}}{\pi\, \hbar}\,. 
\end{equation}
Given the challenge of evaluating gravitational action, the complexity=spacetime volume (CV2.0) proposal \cite{Couch:2016exn} suggests that the complexity is simply determined by the spacetime volume of the WDW patch, viz, 
	\begin{equation}\label{eq:defineCV2}
	\cvv(\Sigma_{\mt{CFT}}) =  \frac{V_\mt{WDW}}{\GN \, \ell^2_{ \rm bulk}}\,. 
\end{equation}

The natural setting where holographic complexity has been investigated is the thermofield double state, which is dual to an eternal two-sided black hole. An appealing feature of the three proposals above is that  they probe  the black hole interior at arbitrarily late times. As evidence that they provide the holographic dual of quantum complexity, the three proposals satisfy two universal features that are argued to hold for any definition of quantum complexity applied to a TFD state.  First, at late times, the complexity should grow linearly with time, with the growth rate proportional to the mass of the dual black hole \cite{Susskind:2014moa}. In addition, there is a universal time delay in the complexity's response to perturbations,  known as the switchback effect \cite{Stanford:2014jda}. 

The various holographic complexity proposals have certainly drawn attention to new types of gravitational observables in the bulk, as well as highlighting their possible role in the AdS/CFT correspondence. However, as described above, the definition of complexity is inherently ambiguous, and so one might expect that the gravitational dual for complexity should also exhibit an analogous and equally diverse set of ambiguities. This idea was realized in recent studies \cite{Belin:2021bga,Belin:2022xmt,Jorstad:2023kmq}, 
where the authors propose a new infinite class of gravitational observables living on general codimension-one or -zero regions of asymptotically AdS spacetimes. It is shown that these new observables exhibit both the expected linear growth at late times and the switchback effect in shockwave geometries. 
Consequently, any member of this infinite class of observables serves as an equally plausible candidate for holographic complexity. This proposal has been somewhat playfully dubbed ``complexity equals anything''.

This note aims to summarize the complexity=anything proposal by exploring several simple examples. We also investigate the properties of the newly introduced gravitational observables and their applications in probing the black hole interiors in the bulk spacetime. 


\section{Complexity Equals Anything}
The complexity=anything proposal was introduced in \cite{Belin:2021bga,Belin:2022xmt}. In part, the impetus behind these studies came from the inherent ambiguities associated with the definition of circuit complexity, and revealed an infinite class of gravitational observables that can be considered plausible candidates for holographic complexity. In particular, these diffeomorphism-invariant observables collectively exhibit linear growth in the late-time regime and manifest the switchback effect with shockwave perturbations, in AdS black hole backgrounds. In this section, we give a succinct overview of the complexity=anything proposal \cite{Belin:2021bga,Belin:2022xmt} with both codimension-one and codimension-zero observables. 

\begin{figure}[t!]
	\centering
	\includegraphics[width=4.5in]{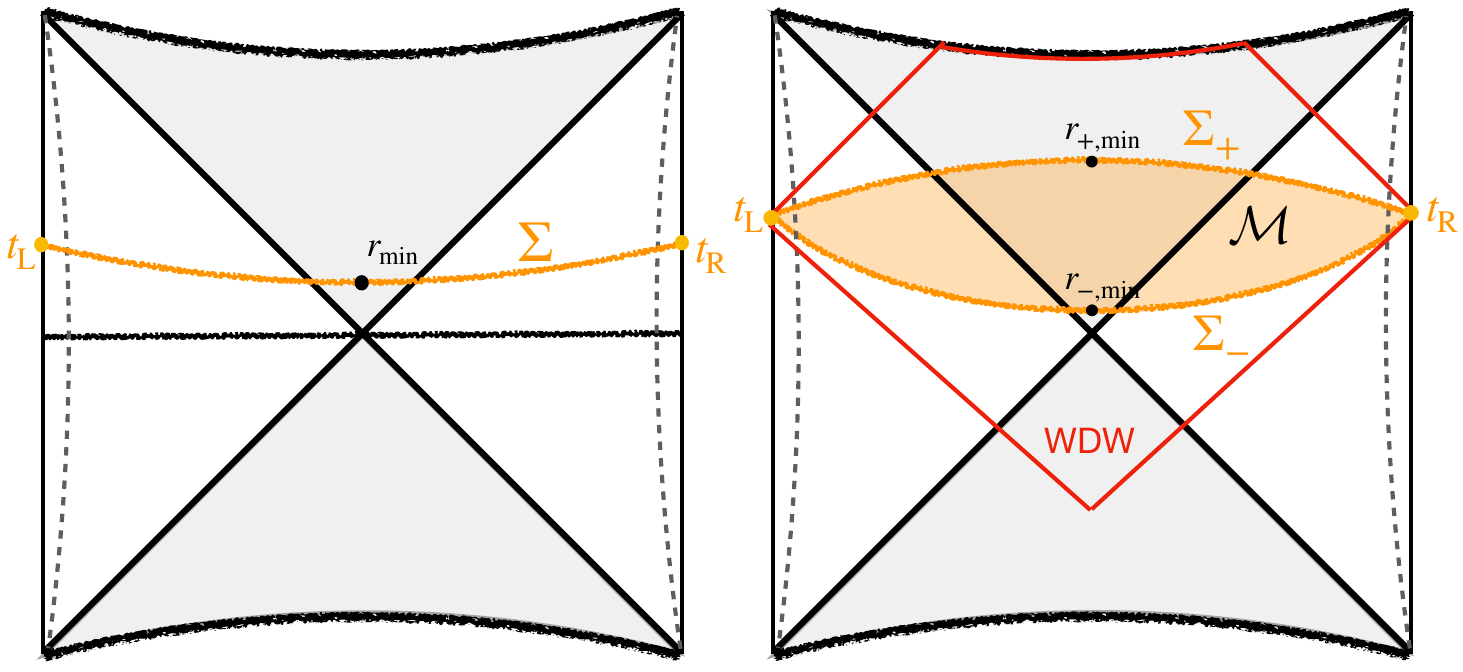}
	\caption{The left panel shows an extremal codimension-one hypersurface {$\Sigma(\tau)$}, represented by the orange curve. On the right, the orange region denotes the codimension-zero subregion $\cal M$ bounded by the two surfaces $\Sigma_\pm$ as in the complexity=anything proposal. In a certain limit, the subregion can extend to the WDW patch with its null boundaries, as shown by the red lines.}\label{fig:TFDbalckhole}
\end{figure}

\subsection{Codimension-one Observables}
Recent investigations of holographic complexity proposals have led to the exploration of novel gravitational observables anchored to time slices $\Sigma_\mt{CFT}$ on the asymptotic boundary. Building upon this discourse, ref.~\cite{Belin:2021bga} extended the CV proposal \eqref{eq:defineCV} to encompass an infinite class of codimension-one observables. An essential observation in this discussion was that the CV proposal \eqref{eq:defineCV} involves a twofold process. First, there is an extremization procedure which identifies a distinctive codimension-one surface, \ie the extremal hypersurface, chosen from the myriad of possible spacelike surfaces with a fixed boundary at $\Scft$. Subsequently, one evaluates a specific geometric property, namely the volume, of this selected hypersurface. Motivated by this perspective, the most general gravitational observables on codimension-one hypersurfaces can be characterized by two scalar functionals, $F_1$ and $F_2$ \cite{Belin:2021bga}. Following the earlier discussion, the first step is to fix a specific codimension-one hypersurface via an extremization procedure. To wit, 
\begin{equation}
	\delta_{\mt{X}} \left[ \int_{\Sigma} d^d\sigma \sqrt{h} \,F_2(g_{\mu\nu};X^\mu)\right] =0 \,,
	\label{stepone}
\end{equation}
where $F_2(g_{\mu\nu};X^\mu)$ depending on the bulk metric $g_{\mu\nu}$ and the embedding functions $X^\mu(\sigma^a)$ specifying the location of $\Sigma$. The extremization is performed over all spacelike hypersurfaces which are asymptotically anchored to a fixed time slice $\Scft$ on the boundary theory -- see the left panel of Fig. \ref{fig:TFDbalckhole}. This procedure selects a particular codimension-one hypersurface, denoted $\Sigma_{F_2}$, and given this hypersurface, the next step is to evaluate a geometric property on the surface to define the gravitational observable,
\begin{equation}\label{eq:obsdef}
	O_{F_1,\Sigma_{F_2}}(\Scft) =\frac{1}{\GN L} \int_{\substack{\Sigma_{F_2} }} d^d\sigma \,\sqrt{h} \,F_1(g_{\mu\nu}; X^{\mu}) \,.
\end{equation}
Again, the scalar functional $F_1$ depends on the bulk metric and the embedding functions. Of course, this definition is simplified with the choice $F_1=F_2$ and the latter reduces to the CV proposal by taking $F_1=F_2=1$. 

To investigate the properties of the gravitational observables defined in eqs.~\reef{stepone} and \eqref{eq:obsdef}, we turn our attention to asymptotically AdS black holes as the ($d$+1)-dimensional background geometry. Employing Eddington–Finkelstein coordinates, the bulk metric takes the following form
\begin{equation}\label{eq:AdSBH}
	d s^{2}=-f(r)\, d v^{2}+2\,dv\,dr+r^{2}\, d{\Omega}^{2}_{k,d-1}\,, 
\end{equation}
where $v=t -\int^\infty_r {d\tilde{r}}/{f(\tilde{r})}$ denotes the infalling time coordinate. The spatial sections of the boundary geometry are spherical, flat or negatively curved with $k=+1,0,-1,$ respectively. The  full two-sided black hole geometry is dual to the TFD state entangling the CFTs on the left and right boundaries, \ie
\begin{equation}\label{eq:TFD}
	\left|\psi_{\mathrm{TFD}}\left(t_{\mathrm{L}}, t_{\mathrm{R}}\right)\right\rangle=\sum_{E_n} e^{-\beta E_n / 2-i E_n\left(t_{\mathrm{L}}+t_{\mathrm{R}}\right) / 2}|E_n\rangle_{\mt{L}} \otimes|E_n\rangle_{\mt{R}} \,.
\end{equation}
Without loss of generality, we consider a symmetric setup for the boundary time slice with $t_{\mt{R}} = t_{\mt{L}} = \tau/2$, as shown in Fig. \ref{fig:TFDbalckhole}. The determination of the extremal surface with respect to the measure $F_2$ functional can be reformulated as a classical mechanics problem with involving the the equation of motion for a non-relativistic particle in an effective potential determined by $F_2$. Importantly, it was shown in \cite{Belin:2021bga} that the growth rate converges to a constant at late times, namely
\begin{equation}
	\lim _{\tau \rightarrow \infty}\left(\frac{d O_{F_1,\Sigma_{F_2}} }{d \tau}\right) \propto P_{\infty} \,.
\end{equation}
Thus one recovers the desired linear growth of these observables at late times. Moreover, the growth rate $P_{\infty}$ is found to be proportional to the mass of the AdS black hole (in the high-temperature limit). At the same time, the evaluation of these observables in the shockwave geometries reveals the manifestation of the switchback effect -- a universal time delay in response to perturbations. 

It is noteworthy that not all of these observables produce extremal surfaces, as described above. As elucidated in \cite{Jorstad:2023kmq}, the existence of the aforementioned in the late-time limit hinges on the presence of a local maximum of the effective potential $U(r)$ within the horizon. Otherwise, there are no locally extremal surfaces and, as described in \cite{Jorstad:2023kmq}, the hypersurfaces $\Sigma_{F_2}$ determined by the maximization process are pushed to the boundary of the phase space of allowed hypersurfaces.

\subsection{Codimension-zero Observables}
The discussion of \cite{Belin:2022xmt} extends these ideas to an infinite class of gravitational observables associated with codimension-zero regions $\mathcal{M}$, as shown in the right panel of Fig.~\ref{fig:TFDbalckhole}.  In particular, the bulk subregion $\mathcal{M}$ is enclosed by the future and past boundaries, denoted $\Sigma_\pm$, both anchored to the same boundary time slice, \ie $\partial\Sigma_\pm=\Scft$. Hence $\partial \mathcal{M}= \Sigma_+ \cup \Sigma_-$. These codimension-zero observables are also shown to exhibit the desired universal properties for holographic complexity, including the linear growth at late times and the switchback effect \cite{Belin:2022xmt}. This construction draws inspiration from the CA and CV2.0 proposals, which evaluate different functionals on the WDW patch. In contrast, however, the complexity=anything proposal begins by identifying a specific bulk subregion through an extremization procedure, akin to the generalized approach in eq.~\eqref{stepone}. 
	
Therefore, the codimension-zero version of complexity=anything again follows a two-step procedure. First, we identify the extremal subregion and boundaries with three independent geometric scalar functionals $F_{2,\pm}$ and $G_2$ by varying the shape of the two boundaries $\Sigma_\pm$, \ie	
\begin{equation}
	\delta_{\mt{X}_\pm}\! \left[  
	\int_{ \Sigma_\pm}\!\!\!  d^d\sigma \sqrt{h} \, F_{2,\pm}(g_{\mu\nu}; X^{\mu}_\pm) 
	+\frac{1}{L } \int_{\mathcal{M}}\!\!\! d^{d+1}x \sqrt{g} \, G_2(g_{
		\mu\nu}) 
	\,\right]=0 \,.
	\label{Xtreme1}
\end{equation}
Here, $F_{2,\pm}$ are scalar functionals of the bulk metric $g_{\mu\nu}$ and the embedding functions $X^\mu_\pm(\sigma^a)$ of the corresponding boundaries $\Sigma_\pm$. Similarly, $G_2$, integrated over the codimension-zero region $\mathcal{M}$, is a scalar functional constructed from the background curvature. This procedure singles out a special codimension-zero region $\mathcal{M}_{G_2,F_{2,\pm}}$ with extremal boundaries  denoted as $\Sigma_\pm[G_2,F_{2,\pm}]$. 
 
 Subsequently, we can evaluate a separate geometric functional on this extremal subregion. The observable is thus defined by 
\begin{equation}\label{eq:O1}
	\begin{split}
		&O\[G_1,F_{1,\pm},\mathcal{M}_{G_2,F_{2,\pm}}\] (\Scft)=\frac{1}{\GN L }\int_{ \Sigma_+[G_2,F_{2,+}]}\!\!\!\!\!\!\!\!\!\!\!\!\!  d^d\sigma \,\sqrt{h} \,F_{1,+}(g_{\mu\nu}; X^{\mu}_+) \\
		&
		+\frac{1}{\GN L }\int_{   \Sigma_-[G_2,F_{2,-}]}\!\!\!\!\!\!\!\!\!\!\!\!\!  d^d\sigma \,\sqrt{h} \,F_{1,-}(g_{\mu\nu}; X^{\mu}_-) +\frac{1}{G_N L^2 } \int_{\mathcal{M}_{G_2,F_{2,\pm}}}\!\!\!\!\!\!\!\!\!\!\!\!  d^{d+1}x \,\sqrt{g} \ G_1(g_{
			\mu\nu})\,.    
	\end{split}
\end{equation}
Importantly, $F_{1,\pm}$ and $G_1$ are three independent scalar functionals, providing the versatility of this definition. The new complexity=anything proposal thus involves two independent bulk scalars, $G_1$ and $G_2$, and four independent boundary scalars, $F_{1,\pm}$ and $F_{2,\pm}$. Notably, this comprehensive proposal subsumes the codimension-one observables originally discussed above by setting $G_1=G_2=F_{1,-}=F_{2,-}=0$, emphasizing its flexibility. 

The simplest non-trivial choice for eq.~\reef{Xtreme1} is obtained by taking $F_{2,\pm}$ and $G_2$ all to be constants. To wit, 
\begin{equation}
	\label{eq:CMCfunc}
	\mC_{\rm gen}=\frac{1}{\GN L } \[\alpha_+ \int_{\Sigma_+}\!\!\!d^d\sigma\,\sqrt{h}+\alpha_- \int_{\Sigma_-}\!\!\!d^d\sigma\,\sqrt{h} + \frac{1}{L}\,\int_{\mathcal{M}}\!\!d^{d+1}x\,\sqrt{-g}\]\,,
\end{equation}
where $\alpha_\pm$ are dimensionless (positive) constants. The result of extremizing over the shape of the boundary surfaces has an elegant geometric interpretation,
\begin{equation}
	\label{eq:CMCextr}
	K_{\Sigma_+} = -\frac{1}{\alpha_+L}\,,\qquad K_{\Sigma_-} =\frac{1}{\alpha_- L}\,. 
\end{equation}
In other words,  the extremal boundaries $\Sigma_\pm$ are constant mean curvature (CMC) slices. An interesting feature of this simple result is that CMC slices can be pushed to the future and past null surfaces emanating from $\Scft$ in the limit $\alpha_\pm \to 0$, as indicated by the divergences $K_{\Sigma_\pm} \to \mp \infty$. That is, $\cal M$ becomes the WDW patch in this limit. So we can recover the CA and CV2.0 proposals as special cases of the complexity=anything proposal in eqs.~\reef{Xtreme1} and \eqref{eq:O1}. 

\section{Singularity Probes}
\begin{figure}[t]
	\centering
	\includegraphics[width=4.5in]{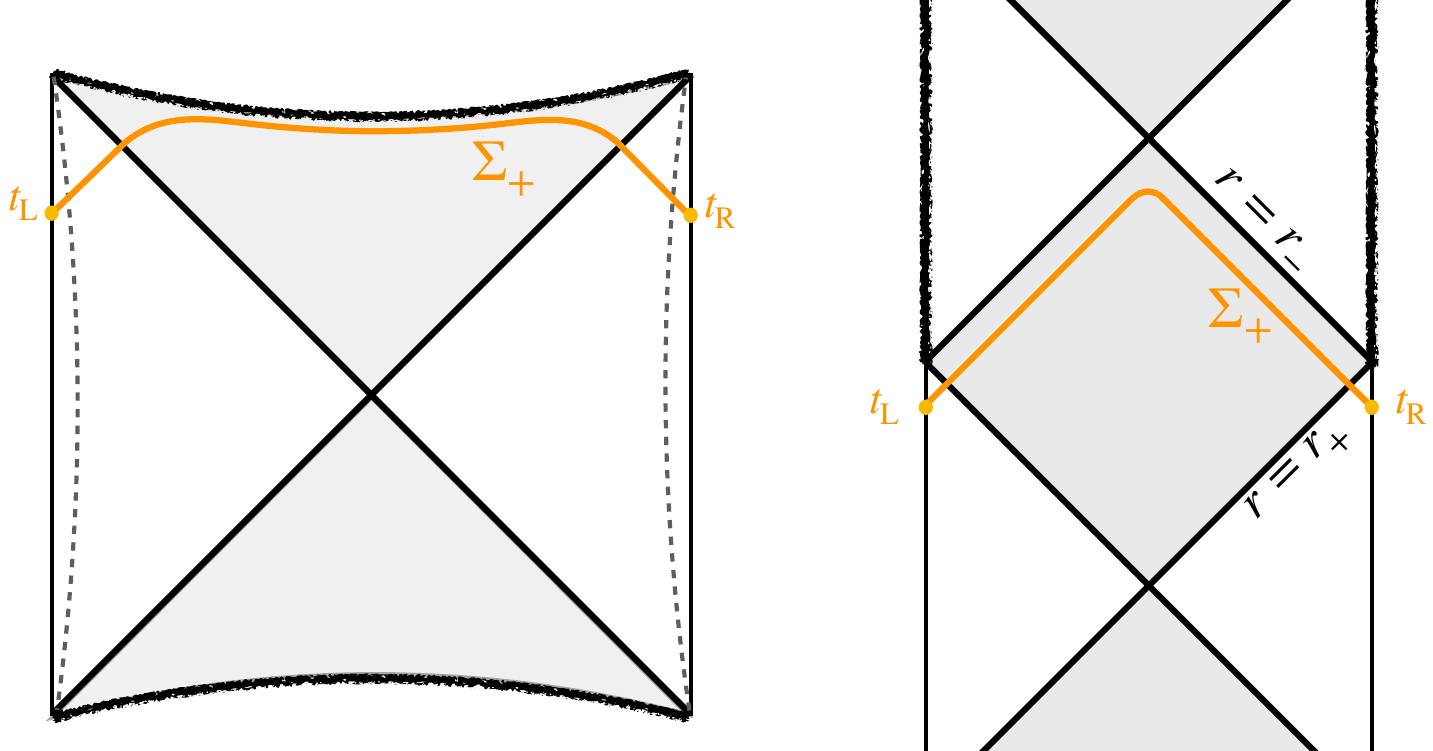}
	\caption{The left and right panels correspond to the Penrose diagrams for the AdS Schwarzschild and AdS Reissner-Nordstr\"{o}m black hole, respectively. Orange curves denote late-time CMC slices $\Sigma_+$ with $\alpha_+\ll 1$. }\label{fig:CAnthingSingularity}
\end{figure}	
Thanks to the flexibility of the complexity=anything proposal, we are able to systematically probe the black hole interior. Furthermore, it provides gravitational observables that facilitate the understanding of the geometric properties near the spacetime singularity \cite{Jorstad:2023kmq}. Of course, the latter requires that we position the extremal surface $\Sigma_\pm$ in close proximity to the singularity. However, this proximity can be achieved by considering the future CMC slice defined in eq.~\eqref{eq:CMCextr} with $\alpha_+ \ll 1$, as shown in Fig. \ref{fig:CAnthingSingularity}. For simplicity, we focus our attention on observables defined by local geometric functionals evaluated only on the future CMC slice $\Sigma_+$. Specifically, the observables \eqref{eq:O1} are evaluated with $F_1=0=G_1$, leaving 
\begin{equation}
	\label{eq:CMCfun}
	\mC_{\mt{CMC}}=\frac{1}{\GN L } \int_{\Sigma_+}\!\!\!d^d\sigma\,\sqrt{h} \, F_{\mt{CMC}}(g_{\mu\nu}, X_+^\mu)\,.
\end{equation} 
The value of the extrinsic curvature \reef{eq:CMCextr} is varied to control the distance between the CMC slice and the singularity at late times. This allows us to probe the geometry near the singularity by studying the late-time growth of these CMC observables. Considering both AdS Schwarzschild and AdS Reissner-Nordstr\"{o}m (RN) black holes, we demonstrate how various properties of the spacetime geometry can be systematically  revealed using this class of gravitational observables. 

We commence our analysis by considering the AdS Schwarzschild black hole, described by the metric in eq.~\eqref{eq:AdSBH} with
\begin{equation}
	f(r)=k+\frac{r^2}{L^2}-\frac{\omega^{d-2}}{r^{d-2}} \,,
\end{equation}
where the parameter $\omega$ determines the mass of the AdS black hole. At late times, the extremal surface $\Sigma_+$ approaches a constant radial slice, denoted by $r=r_f$. In the limit $\alpha_+ \ll 1$, the CMC slice approaches the future singularity with
\begin{equation}
	\lim\limits_{\alpha_+ \to 0}	r_f \simeq   \( \frac{d^2L^2 \omega^{d-2}}{ 4 }  \)^{1/d} \, \alpha_+^{2/d}  \,.
\end{equation}
Given this surface, we evaluate the CMC observable \eqref{eq:CMCfun} with different choices of the geometric functional $F_{\mt{CMC}}$. Specifically, we consider:
\begin{equation}\label{eq:threeF}
	F_{\mt{CMC}} =1\,, \qquad F_{\mt{CMC}}= - L\,K\,, \qquad  F_{\mt{CMC}}= L^4C^2 \,, 
\end{equation}
where as above, $K$ denotes the trace of the extrinsic curvature  and $C^2= C_{\mu\nu\rho\sigma}C^{\mu\nu\rho\sigma}$ is the square of the Weyl tensor. 
The growth rate at late times associated with these three observables can be derived as 
\begin{equation}
	\lim\limits_{\tau  \to \infty} \frac{d 	\mC_{\mt{CMC}}}{ d \tau }  \simeq
	\begin{cases}
		\frac{8\pi d }{(d-1)}\,M\, \alpha_+\,,& F_{\mt{CMC}} =1 \,,\\
		\,\\
		\frac{8\pi d }{(d-1)}\,M \,,&  F_{\mt{CMC}}= - L\,K\,,\\
		\,\\
		\frac{128 \pi (d-1)(d-2)}{d^2}\, \frac{M}{\alpha_+^3}  \,,& F_{\mt{CMC}}= L^4C^2\,.\\
	\end{cases}
\end{equation}
In summary, the above results reveal different behaviours for the various observables as the CMC slice approaches the singularity. Of particular note is the unbounded growth rate (with $1/\alpha_+^3$) in the case of $L^4C^2$, providing evidence that the extremal surface $\Sigma_+$ is close to a curvature singularity. For comparison purposes with the following recall that $\frac{d}{d-1}\,M = S\,T$ where $S$ and $T$ denote the entropy and temperature of the black hole, respectively.

Having investigated the behaviour of various CMC observables near the singularity of the AdS Schwarzschild black hole, we contrast these results with those for the AdS Reissner-Nordstr\"{o}m black hole -- see the right panel of Fig. \ref{fig:CAnthingSingularity}. The metric for the latter is still expressed by eq.~\eqref{eq:AdSBH}, where the blackening factor is instead given by
\begin{equation}
	f_{\mt{RN}}(r)=k+\frac{r^2}{L^2}-\frac{\omega^{d-2}}{r^{d-2}}+\frac{q^2}{r^{2(d-2)}} \,. 
\end{equation}
Unlike the AdS Schwarzschild geometry, the AdS RN black hole features a timelike singularity and inner/outer horizons at $r=r_\pm$, where $f(r_\pm)=0$. Consequently, the timelike singularity remains beyond the region probed by the CMC slices anchored to the asymptotic boundaries. The CMC slices effectively probe the black hole interior only up to the inner horizon located at $r=r_-$. It proves advantageous to express the resulting quantities in terms of the Bekenstein-Hawking entropy and Hawking temperature associated with the inner horizon, 
\begin{equation}
	S_{-}=\frac{\Omega_{k, d-1} r_{-}^{d-1}}{4 G_{\mt{N}}} \quad \text { and } \quad T_{-}=\frac{\left|f_{\mt{RN}}^{\prime}\left(r_{-}\right)\right|}{4 \pi} \,.
\end{equation}
For example, in the regime of large extrinsic curvature (\ie $\alpha_+\ll1$), the position of the final slice $r=r_f$ is approximately located at
\begin{equation}
	r_f \simeq r_{-}  + 4\pi L^2 T_- \,\alpha_+^2  \quad \text {with } \quad  \alpha_+ \ll 1\,. 
\end{equation}
Similar to the previous analysis, the late-time growth rate for the three observables \eqref{eq:threeF} can be expressed as
\begin{equation}
	\lim\limits_{\tau  \to \infty} \frac{d 	\mC_{\mt{CMC}}}{ d \tau }  \simeq
	\begin{cases}
		8 \pi S_- T_- \,\alpha_+  \to 0\,,& F_{\mt{CMC}} =1 \,,\\
		8 \pi S_- T_-  \,,&  F_{\mt{CMC}}= - L\,K\, ,\\
		\# S_- T_- \,\alpha_+ \to 0  \,,& F_{\mt{CMC}}= L^4C^2 \,,\\
	\end{cases}
\end{equation}
where $\#$ in the last line is a complicated function of $\omega$, $q$ and $k$.
The vanishing of the growth rate for $F_{\mt{CMC}} =1$ or $ C^2 L^4$ can be attributed to the vanishing of the volume element as the CMC slice approaches a null surface (\ie the inner horizon), while the Weyl-squared term remains finite there. 
\section{Summary and Outlook}
In summary, our brief review has delved into the recent ``complexity equals anything" proposal \cite{Belin:2021bga,Belin:2022xmt,Jorstad:2023kmq} for holographic complexity. These investigations have revealed an infinite class of gravitational observables defined for codimension-one or -zero subregions of the bulk spacetime. Somewhat surprisingly, (almost) each of these holographic observables stands as an equally plausible candidate for the gravitational dual of the circuit complexity of the corresponding boundary state. A key lesson is that the freedom in constructing gravitational observables is an innate feature  of holographic complexity, rather than a shortcoming. As we illustrated, this freedom leads to families of gravitational observables which allow for a systematic study of the black hole interior in a way that would be impossible with, \eg the CV or CA proposals alone. Furthermore, it seems to mimic the ambiguities found in the definition of circuit complexity in the boundary theory.  Of course, one of the interesting future research directions would be to determine a more explicit relation between the ambiguities found in the bulk and the boundary constructions. This would undoubtedly lead to progress in understanding the precise interpretation of the gravitational observables in terms of the observables constructed in the boundary theory. Undoubtedly, this new proposal for holographic complexity points to numerous new opportunities for research at the intersection of quantum gravity and quantum information.

	
	\section*{Acknowledgments}	
RM presented this material at {\it Gravity, Strings and Fields: A conference in honour of Gordon Semenoff}, held July 24-28, 2023 at the Centre de Recherches Math\' ematiques. We would like to wish Gordon a happy birthday and to congratulate him on his enormous successes and contributions to theoretical physics.
	We would also like to thank Alex Belin, Eivind J\o rstad, Gabor S\' arosi and Antony Speranza for enjoyable collaborations on the material discussed here. Research at Perimeter Institute is supported in part by the Government of Canada through the Department of Innovation, Science and Economic Development Canada and by the Province of Ontario through the Ministry of Colleges and Universities. RCM is supported in part by a Discovery Grant from the NSERC of Canada, and by the BMO Financial Group. RCM and SMR are supported by the Simons Foundation through the “It from Qubit” collaboration. SMR is also supported by MEXT-JSPS Grant-in-Aid for Transformative Research Areas (A) “Extreme Universe”, No. 21H05187 and by JSPS KAKENHI Research Activity Start-up Grant Number JP22K20370.

	%
	%
	
\end{document}